\documentclass[sigconf]{acmart}

\AtBeginDocument{%
  }

\copyrightyear{2023} 
\acmYear{2023} 
\setcopyright{iw3c2w3}
\acmConference[SIGIR '23]{Proceedings of the 46th International ACM SIGIR Conference on Research and Development in Information Retrieval}{July 23--27, 2023}{Taipei, Taiwan}
\acmBooktitle{Proceedings of the 46th International ACM SIGIR Conference on Research and Development in Information Retrieval (SIGIR '23), July 23--27, 2023, Taipei, Taiwan}\acmDOI{10.1145/3539618.3591931}
\acmISBN{978-1-4503-9408-6/23/07}

\usepackage{booktabs} 
\usepackage{xspace}
\newcommand{\ie}{\emph{i.e.,}\xspace}
\newcommand{\eg}{\emph{e.g.,}\xspace}

\begin{document}

\title{Take a Fresh Look at Recommender Systems from an Evaluation Standpoint}

\author{Aixin Sun}
\orcid{0000-0003-0764-4258}
\affiliation{%
  \institution{School of Computer Science and Engineering\\ Nanyang Technological University}
  \streetaddress{Nanyang Avenue}
  \country{Singapore}
  \postcode{639798}
}
\email{axsun@ntu.edu.sg}


\begin{abstract}
 Recommendation has become a prominent area of research in the field of Information Retrieval (IR). Evaluation is also a traditional research topic in this community. Motivated by a few counter-intuitive observations reported in recent studies, this perspectives paper takes a fresh look at recommender systems from an evaluation standpoint. Rather than examining metrics like recall, hit rate, or NDCG, or perspectives like novelty and diversity, the key focus here is on \textit{how these metrics are calculated when evaluating a recommender algorithm}. Specifically, the commonly used train/test data splits and their consequences are re-examined. We begin by examining common data splitting methods, such as random split or leave-one-out, and discuss why the popularity baseline is poorly defined under such splits. We then move on to explore the two implications of neglecting a global timeline during evaluation: \textit{data leakage} and \textit{oversimplification of user preference modeling}. Afterwards, we present new perspectives on recommender systems, including techniques for evaluating algorithm performance that more accurately reflect real-world scenarios, and possible approaches to consider decision contexts in  user preference modeling.
\end{abstract}

\ccsdesc[500]{Information systems~Recommender systems}
\ccsdesc[300]{Information systems~Collaborative filtering}

\keywords{Recommendation, global timeline, practical evaluation, user preference modeling}

\maketitle


\section{Introduction}
Out of all the papers published in SIGIR 2022, 27.5\% of them have titles that include the words ``recommender'' or ``recommendation''.\footnote{\url{https://dblp.org/db/conf/sigir/sigir2022.html}} This is a strong indication of research interests on Recommender Systems (RecSys) in the Information Retrieval (IR) community. As evaluation is also a traditional research topic in IR, it is interesting to study how recommendation algorithms are evaluated in general. More interestingly, a few recent papers report counter-intuitive observations made from experiments on recommender system, both in offline and online settings~\cite{ICTIR22Loyal,EMarket22consumer,BDR22Fairness,IPM21Venue,Perspective22forgetting}. 

Here are some example counter-intuitive observations. \citet{ICTIR22Loyal} report that both users who spend more time and users who have many interactions with a recommendation system receive poorer recommendations, compared to users who spend less time or who have relatively fewer interactions with the system. This observation holds on recommendation results by multiple models (\ie BPR~\cite{BPR}, Neural MF~\cite{neuMF}, LightGCN~\cite{lightgcn}, SASRec~\cite{SASRec} and TiSASRec~\cite{timeSasRec}) on multiple datasets including MovieLens-25M, Yelp, Amazon-music, and Amazon-electronic. On  a large Internet footwear vendor, through online experiments, \citet{EMarket22consumer} observe  that  ``experience with the vendor showed a negative correlation with recommendation performance''. The factors considered under ``experience'' include the number of days since account creation, number of days since the first shopping transaction, and the number and the value of purchase transactions made in the past year. Another study reports that  ``using only the more recent parts of a dataset can drastically improve the performance of a recommendation system''~\cite{Perspective22forgetting}.

\begin{table*}[t]
    \centering
      \caption{The 5 offline evaluation settings described in~\cite{Gunawardana2022}, from the ideal and most close simulation of online process (Setting 1) to the least (Setting 5). The last column indicates whether the data split observes global timeline in user actions.}
    \label{tab:5settings}
    \begin{tabular}{c|p{5.3in}|c}
    \toprule
      Setting   &  Train/test data split scheme & Global timeline\\ \midrule
1 & Step through user actions in temporal order and make predictions for each user action along the way, based on the known user actions at the prediction time. Before the testing time point, every user action serves as a test instance, and subsequently becomes a training instance.& Yes\\ \midrule
2 & Following Setting 1, instead of evaluating all user actions along time, only evaluate sampled user actions as test instances. The only difference to Setting 1 is the reduced  number of test instances along the way. & Yes\\\midrule
3 & Sample a set of test users, then sample \textit{a single test time}, and hide all items of test users after that time point. That is, the data is partitioned to train/test sets based on a single time point. & Partially\\ \midrule
4 & Sample a test time for each test user (\eg right before user's last action), and do not observe global timeline across all users. Leave-one-out is an example data split scheme under this setting. & No\\  \midrule
5 & Completely ignore time as in the case that timestamps of user actions are unknown. Data is randomly partitioned into train and test sets. & No \\ 
    \bottomrule 
    \end{tabular}
  
\end{table*}

We  interpret the reported counter-intuitive observations from two perspectives. First, these observations are made with respect to the time dimension, more specifically, the \textbf{global timeline} of  user-item interactions. Here, we are not considering time as an additional feature or context in the algorithm modeling. Rather, we consider the arrangement of the user-item interactions by their timestamps in chronological order during evaluation.\footnote{Although there are recommendation models which consider time as a contextual feature in their modeling, not many studies arrange user-item interactions along the global timeline chronologically, and consider the \textit{absolute time points} of the interactions in their evaluations. We will use a case study to support this claim shortly.}   Hence, we have  ``number of days since the first transaction'' and ``recent parts of a dataset''. The reported counter-intuitive observations call for a revisit of \textit{the importance of observing the global timeline in evaluating recommender models}. Findings from the revisit may impact our way of conducting evaluation, in turn the model design, and more importantly, our understanding of recommender system. Second, these observations are considered counter-intuitive because they contradict our expectation (or an implicit assumption) on a recommender. That is, the more interactions a user has with a system, the higher chance that the recommender better learns the user's preference. However, these observations show otherwise.  

With global timeline in mind, we conduct a case study to find out: to what extent the global timeline is observed in offline evaluation in academic papers (Section~\ref{sec:dataSplit}). Our case study is based on the full and industry papers published in the ACM Recommender Systems conference in the past three years (2020, 2021, 2022). Based on the findings, we revisit Popularity, the simplest recommendation model, in Section~\ref{sec:popularity} to justify why this commonly used baseline is ill-defined. Then we move on to the discussion on the consequences of ignoring global timeline in evaluation: data leakage (Section~\ref{sec:dataLeakage}) and simplification of user preference modeling (Section~\ref{sec:simpliPerference}). In Section~\ref{sec:freshlook}, we propose a fresh look at recommender system from the evaluation perspective.  In Section~\ref{sec:discussion}, we present a summary of the key messages and contributions of this work, after which the paper is concluded in Section~\ref{sec:conclude}.

\section{Case Study: Data Split Schemes}
\label{sec:dataSplit}

Most academic researchers do not have access to an online platform to directly evaluate their models by real user-item interactions. Evaluation on an offline dataset is the only choice in most cases. It is also well known that there are many more factors that may affect user behaviour online and the prediction power collected from offline evaluations may or may not be observed online. Hence, ``the goal of the offline experiments is to filter out inappropriate approaches, leaving a relatively small set of candidate algorithms to be tested'' online, as stated in the evaluation chapter of the recommender systems handbook~\cite{Gunawardana2022,RecSysHB2022rsh}.  However, to conduct offline evaluation, ``it is necessary to simulate the online process where the system makes predictions or recommendations''~\cite{Gunawardana2022}. Apparently, a close simulation of online process would make the results obtained from offline evaluation more indicative, better serving the purpose of algorithm selection.

Table~\ref{tab:5settings}  summarizes the five settings described in~\citet{Gunawardana2022} from the ideal setting (Setting 1) of simulating the online process as close as possible, to the most simplified setting (Setting 5). For simplicity, in our discussion we only consider training and test instances, and do not consider validation or development set. We remark that the last two settings (Settings 4 and 5) do not maintain or observe global timeline across all users. Hence, these two settings are not considered as close simulations of the online recommendation processes. As for Setting 3, the partition of train/test sets is based on a single time point along the global timeline. However, within the train or test sets, the data instances may not maintain their temporal order. 

To understand which settings are more widely used in evaluating recommender systems, we conducted a case study to collect the data split schemes used in the papers published in the last three years (2020 - 2022) of ACM Recommender Systems conference. The ACM RecSys conference is considered here for its strong relevance to the topic and reasonable size. We considered all full papers and industry papers. However, a good number of papers study recommenders from system perspective like training efficiency, distributed and/or federated RecSys. Some others focus on user studies and user preference analysis. Hence, we did not include these papers in the case study. After filtering, we had 82 full and 9 industry papers which had clear descriptions of experiment settings. Among them, we further excluded another 3 papers. Two of them design experiments dedicated to cold-start setting, and one paper is on news recommendation and the data is split by news topic in their experiment. Finally, our case study included 88 papers.

\begin{table}[t]
    \centering
     \caption{Number and percentage of papers by their adopted data split scheme in ACM RecSys conference (2020 - 2022). 
    \label{tab:datasplit}}
    \begin{tabular}{rr|l|c}
    \toprule
    \multicolumn{2}{c|}{No. and \% papers} & Data split & Global timeline\\
    \midrule
      30 & 34.1\% & Random split & No \\
      22 & 25.0\% & Leave-one-out & No\\
      17 & 19.5\% & Single time point & Partially\\
      15 & 17.0\%& Simulation-based online & Yes\\
    4&4.5\%    & Sliding window & Yes\\ 
    \bottomrule
    \end{tabular}
   
\end{table}

Summarized in Table~\ref{tab:datasplit}, out of the 88 papers, 30 papers adopt random split (\ie Setting 5). The random split can be either global-based or user-based. For the latter, a percentage of actions from each user are randomly sampled as test instances; then the remaining are training instances. The next popular split scheme with 22 papers is leave-one-out (\ie Setting 4), also known as leave-last-one-out, where the last action of each user is a test instance. We also include the cases where the last few actions (\eg based on a pre-defined ratio) of each user are test instances. Then, 17 papers split data by a single time point (\ie Setting 3), and 15 papers utilize simulation-based online testing. We note that the latter 15 papers mainly focus on Bandits and reinforcement learning for recommendation. Lastly, there are 4 papers using sliding window for evaluating incremental learning or session based learning. 

From this case study, we understand that 59.1\% of the collected papers follow Settings 4 and 5 (see Table~\ref{tab:5settings}). That is, their offline evaluation do not well simulate the online process, and global timeline across users is not maintained. We also observe that, despite 17\% of the papers utilizing a simulation-based online setting, the reason for the adoption is not due to the requirements of the recommendation research problem, but rather due to the algorithms employed in their solutions, \eg reinforcement learning. Discussion on the evaluation of reinforcement learning-based recommenders~\cite{reinforcementEve23} is beyond the scope of this paper.

We also note that, there are other data split schemes for evaluating recommenders~\cite{rigorousEvaluation,exploreDataSplit,futureItemRec}. Our  focus here is whether global timeline is maintained during the offline evaluation. Next, we revisit the Popularity baseline to illustrate why maintaining global timeline is critical.

\section{Ill-defined Popularity}
\label{sec:popularity}
As a non-personalized method, Popularity is often considered as the simplest  baseline and is widely used for comparison purpose in evaluation. Specifically, in our case study, 26\% of the papers use Popularity as one of their baselines. In academic  papers and the toolkits for RecSys, the popularity of an item is mostly defined as the number of interactions it receives in the training set. Recall that in our case study, 59.1\% of the papers do not observe global timeline in their evaluation, hence they cannot define popularity of items along time. Then an interesting question here: \textit{To what degree does the popularity, as determined by the frequency of an item in the training set, reflect its true popularity in a real-world scenario?}

\begin{figure}[t]
    \centering
	  \includegraphics[trim = 4.1cm 13.4cm 15.1cm 0.2cm, clip, width=0.95\columnwidth]{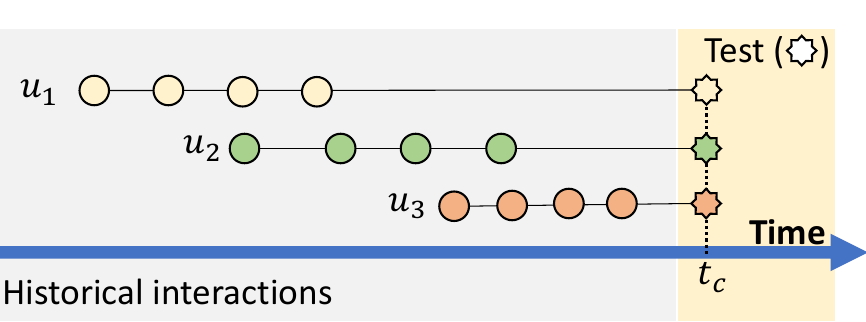}
    	\caption{Train/test in practical systems, where $t_c$ indicates the current time point.}
        \label{fig:timePractice}
	\Description{}
 \end{figure}

To answer this question, we first show how a recommender works in general. With the help of a global timeline, Figure~\ref{fig:timePractice} gives absolute time points of historical user-item interactions for three example users $u_1$, $u_2$, and $u_3$. In the illustration, $t_c$ indicates the current time point. If the users visit the website at the current time $t_c$, then the system will make recommendations to these users. Users then may choose to interact or not to interact with the recommended items. In the illustration, let us assume that all users interact with the recommended items, and these interacted items become the latest interactions for all  three users. In practice, a recommender can learn from all or a subset of historical interactions that occurred before time $t_c$, and makes recommendations if users visit the site at time $t_c$.

\begin{figure}[t]
    \centering
    \includegraphics[width=0.95\columnwidth]{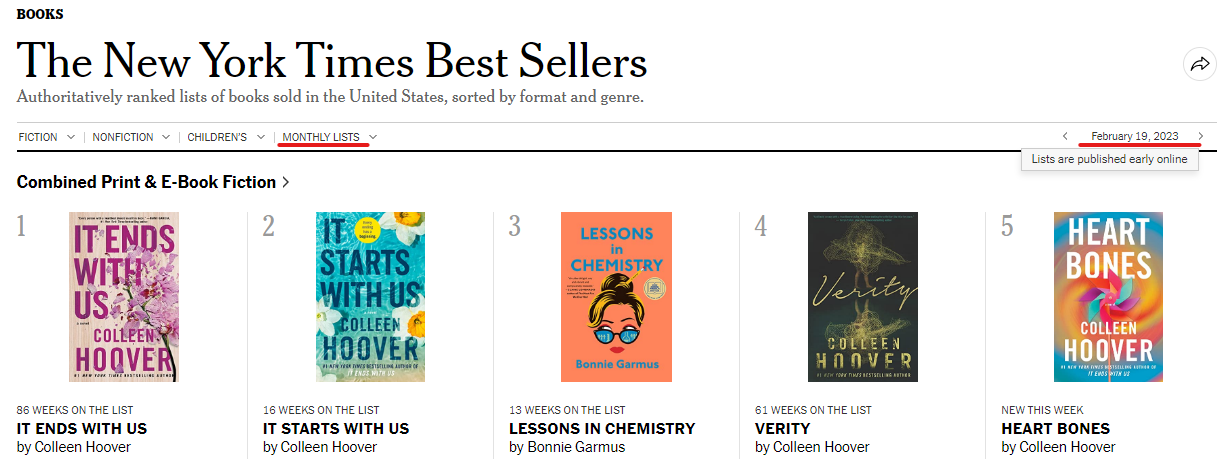}
    \caption{The New York Times best sellers, 19 Feb 2023}
    \label{fig:NYpopular}
\end{figure}

Now, let us look at some real-world examples of popularity ranking. The New York Times best sellers\footnote{\url{https://www.nytimes.com/books/best-sellers/}} is a well-known and influential list of best-selling books in the United States. The list is updated on a \textit{weekly basis} since 1931\footnote{\url{https://en.wikipedia.org/wiki/The_New_York_Times_Best_Seller_list}} and Figure~\ref{fig:NYpopular} shows a screencapture of the top few books as at the week of 19 Feb 2023.  There is also an option to display \textit{monthly} lists of different book genres, as shown in the left hand side of Figure~\ref{fig:NYpopular}. The best sellers on Amazon is an hourly updated list.\footnote{\url{https://www.amazon.com/bestsellers}}  In short, it is typical for websites to feature a popularity ranking of items that is updated on an hourly, daily, weekly, or monthly schedule. 

The real-world popularity rankings have two important properties. First, the rankings are dynamically updated along timeline. For instance, the best sellers on the New York Times in Week 1 of Year 2020 shall be quite different from that in Week 3 of Year 2023. Second, the popularity ranking only considers item frequency in a predefined time range (\eg an hour, a week, or a month), and does not necessarily use all historical data. The current weekly ranking only needs to use the interactions occurred  in the past week, and the current hourly ranking only use interactions in the past hour. In other words, popularity has  a very strong transience effect and often refers what are trending during a (relatively short) time period.

The popularity baseline widely used in RecSys offline evaluations is very different. There is no time window (\eg a week or a month) defined, and the popularity ranking is not updated along timeline. In fact, as a significant portion of papers (\eg 59.1\% in Table~\ref{tab:datasplit}) do not maintain global timeline of their user actions, it is not possible to define and update such a ranking along timeline. Hence, the popularity baseline is ``forced'' to use all interactions in training set. As the result, the frequency-based ranking is a \textit{static} ranking, covering the entire duration of the training data. This duration is determined by the dataset, and also by the adopted data partition scheme. For instance, the duration of popularity for leave-one-out scheme will be the duration of the entire dataset, and the duration for the single time point scheme would be defined by the data before the time point.

If the data points in a dataset indeed cover a short time period, then popularity remains indicative. However, if the dataset covers user interactions collected from a long time period, then a single static ranking completely ignores the transience of popularity. This ranking becomes less meaningful. In particular, many datasets like MovieLens, Yelp, and Amazon reviews cover data points in a very long time span, \eg more than 10 years~\cite{dataLeakage}. In such datasets, a single static popularity ranking will be very different from the kind of ranking we see in Figure~\ref{fig:NYpopular}. \citet{popularity20} report that if we follow the popularity definition in real world, the performance of the popularity method increases by at least 70\%, compared to the ill-defined popularity on MovieLens dataset.

\section{Data Leakage}
\label{sec:dataLeakage}

\begin{figure}[t]
    	\centering
    	\includegraphics[trim = 4.1cm 7.5cm 15.1cm 6.8cm, clip, width=0.95\columnwidth]{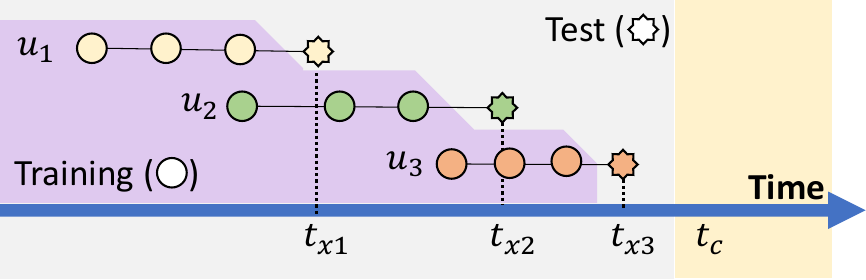}
    	\caption{Train/test with leave-one-out split}
        \label{fig:timeLeaveOneOut}
	\Description{}
\end{figure}

The ignorance of transience of popularity is not the only issue for not observing global timeline. Another major consequence is data leakage, or more specifically accessing future data that is impossible to access in reality. 

In the following, we again use Popularity to illustrate the issue by mapping the data instances onto a timeline. Recall that the two most widely adopted data split schemes are random split and leave-one-out (see Table~\ref{tab:datasplit}). In our following discussion, we use leave-one-out in our examples, to avoid potential confusions of different partitions by random. Leave-one-out, or leave-last-one-out, is to sample the last interaction of each user as a test instance. The user's remaining interactions are in the training set. 

Figure~\ref{fig:timeLeaveOneOut} provides an  illustration of the leave-one-out data partition for three example users, with respect to the global timeline. Observe that the test instance for $u_1$ occurs at time $t_{x1}$. If we consider time $t_{x1}$ to be the current time $t_c$ as if the offline evaluation were online, then all the historical interactions a recommender can learn from at $t_{x1}$ should be the three interactions by $u_1$ and the one interaction by $u_2$. A recommender would never have access to the future interactions that \textit{will happen in the future} with respect to time point $t_{x1}$. At $t_{x1}$, the future items include two interactions by $u_2$ and  all interactions by user $u_3$, which occur after $t_{x1}$. By forcing the  popularity baseline to use \textit{all training data}, the popularity method may recommend some items to $u_1$ that are very popular in future, with respect to time $t_{x1}$. Clearly, recommending items that are popular in future by using the frequency counting that happened in future is unrealistic.

\begin{figure}
    \centering
    \includegraphics[width=0.9\columnwidth]{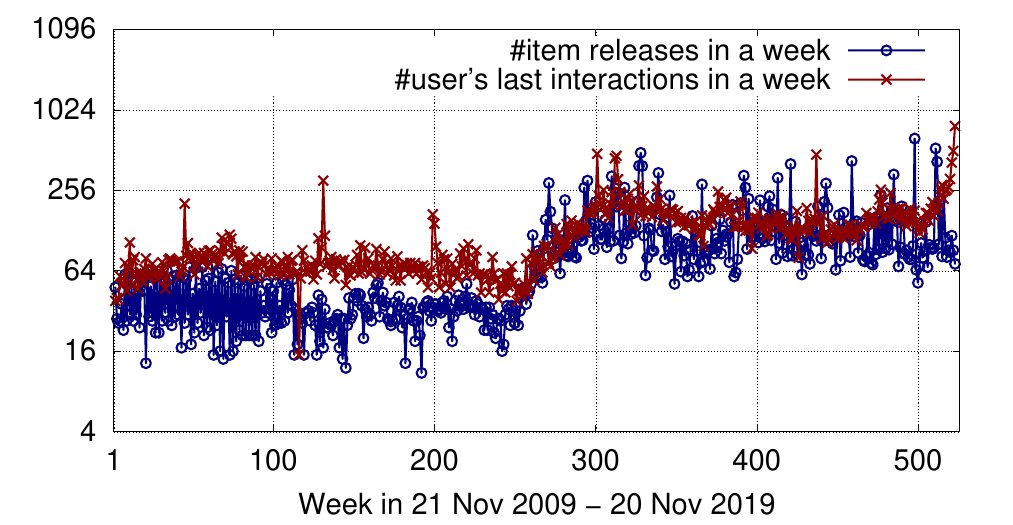}
    \caption{Number users whose last interaction occurred, and the number of movies which receives its very first rating (\ie item release) in each week, in the 10 years (or 520 weeks) period  of the MovieLens-25 dataset    
    (reproduced from~\cite{dataLeakage}).}
    \label{fig:movielens}
\end{figure}

While Figure~\ref{fig:timeLeaveOneOut} is an illustration, we are more concerned about the scenarios that occur in reality. In reality, users may interact with a system at any time; new items may become available in the system at any time; outdated items are removed from the system at any time.  For example, an iPhone model is usually discontinued after two years of its release, and the phone changes from its first generation in 2007 to iPhone 14 in 2022. Many widely used datasets in RecSys research cover user-item interactions collected in a long time period, \eg more than 10 years for MovieLens, Yelp, and Amazon.  Figure~\ref{fig:movielens} plots the number of users whose last interaction occurred in each week, and the number of new movie releases in each week, in 10 years (or 520 weeks) time period in the MovieLens-25M dataset~\cite{dataLeakage}.\footnote{Note that, MovieLens is a rating dataset. The timestamp of rating a movie does not reflect the  time of interacting (\eg watching) the movie. However, we may safely assume that a rating comes after the user has interacted with the movie. We consider the time when a movie receives its very first rating from any user its release time.}  Recall that user last interaction also indicates the test time point like  $t_{x1}$ and $t_{x2}$ in Figure~\ref{fig:timeLeaveOneOut}.

This leads to the next question: \textit{Why the popularity baseline in academic research is evaluated in this way?} The reason is simple. We want to  ensure a ``fair comparison'', where all models are expected to learn from  the same training set, and to be evaluated on the same test set. Basically, our machine learning- or deep learning-based models are trained on the training set and evaluated on the test set. The popularity baseline is treated as a trivial machine learning model. It takes in all instances in the training set and produces a ranking by the ill-defined popularity for the purpose of ``fair comparison'' with others. In fact, it is not difficult to simulate a popularity ranking along timeline with scheduled updates, as a non-personalized recommendation method. However, due to the presumed requirement of ``fair comparison'', all the training and test instances are processed in the same way in an experimental comparison.

Unfortunately, due to the leave-one-out data partition scheme, all machine learning-based or deep learning-based models suffer from the same issue: accessing future data that is impossible to access in reality. Formally, this is known as data leakage in machine learning. \citet{dataLeakage} offer a detailed study on this topic in the context of RecSys, and~\citet{futureItemRec} also observe data leakage in their evaluation.  If a dataset covers a long time period, the items (\eg movie, product, restaurant) are not all available for interaction at the very beginning of the entire time period, and users' last interactions  may occur at any time points. Hence, data leakage is unavoidable if random or leave-one-out data splits are adopted on such datasets. From this perspective, both the simple model like popularity and the more complex models may not be evaluated in a practical manner because they access future data, unless the data partition scheme leads to no (or at least minimum) data leakage. For example, if the dataset is partitioned by an absolute time point (\ie all interactions occurred before $t_s$ are training data, after are test data), then there is no data leakage. Table~\ref{tab:datasplit} shows fewer than 20\% of papers use single time point split.

In summary, the evaluations conducted without observing the global timeline in an offline setting may suffer from data leakage, rendering to incomparable results~\cite{dataLeakage}. Experiments conducted in this way also contribute to the difficulty of reproducibility~\cite{futureItemRec}. In fact, a simple way of partitioning data into train/test without considering timeline is also a form of simplification to the  RecSys research problem.  With time taken into consideration in recent evaluations~\cite{ICTIR22Loyal,CVTT,Perspective22forgetting,StreamingSession22}, we start to obtain counter-intuitive observations listed in the Introduction section.

\section{Simplified User Preference Modeling}
\label{sec:simpliPerference}

The discussion so far on the ignorance of global timeline does not explain (i) \textit{why models trained by using only the more recent parts of data demonstrate better performance?} and (ii) \textit{why more interactions from users lead to poorer recommendations?} A hypothetical answer to both questions is \textit{the simplification in learning user preference} in current models. The missing of the global timeline could be a contributing factor as well. 

To better understand user preference modeling, let us go back 30 years to learn how collaborative filtering was firstly defined in the Tapestry system~\cite{CF92History}. The way that Tapestry system supports collaborative filtering is to let users to read documents recommended by other users. The authors give an example in their original paper~\cite{CF92History}. A user $u$ wants to read interesting but not all documents from a newsgroup. She knows that some users read all of these documents and mark the interesting ones. She then can simply choose to read only the documents that are marked interesting by these users. Such kind of filtering is conceptually similar to reading only the tweets written or retweeted by the users one follows on Twitter.

In Tapestry, user preference is directly reflected by the other users he/she follows. A hypothetical extension of the understanding is that if user $u_1$ follows $u_2$, then $u_1$ prefers $u_2$'s \textit{decision making} in judging interesting documents (or retweeting) given the context at that time, \eg when a document is received in the newsgroup.

Different from Tapestry, where users are empowered to choose who to follow,  user preference in mainstream RecSys research is \textit{inferred} from user-item interactions. The main underlying assumption is that a user $u$ would prefer the items that are chosen by other users who share similar preferences with $u$. Preference similarity between users is reflected by similar user-item interactions in the past. If users $u_1$ and $u_2$ both purchased the same mobile phone, then we would consider that $u_1$ and $u_2$ share similar preference, at least on this particular item. However, purchasing the same phone may not necessarily reflect that the two users share a similar \textit{decision making} process, if we consider the context changes in a system from time to time. 

\begin{figure}[t]
    	\centering
    \includegraphics[trim = 4.1cm 1.5cm 15.1cm 12.3cm, clip, width=0.95\columnwidth]{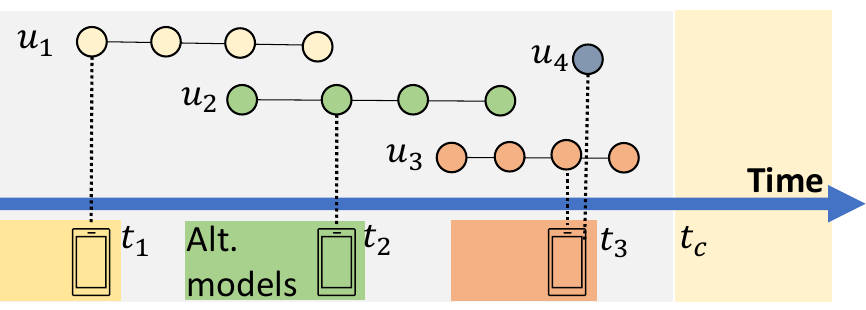}
    	\caption{Context for decision making of an interaction \eg phone purchase}
        \label{fig:timeDecisionContext}
	\Description{}
    \end{figure}

Figure~\ref{fig:timeDecisionContext} shows an example scenario where the three users $u_1$, $u_2$, and $u_3$ purchased the same phone at different time points, $t_1$, $t_2$, and $t_3$ respectively.  We may further consider that $t_1$ is the first day when this phone was released, and $t_3$ is among the last few days when this phone was to be discontinued, and $t_2$ is the middle time in between. We may also consider that an upgraded version of this phone has been released in between $t_2$ and $t_3$. In this scenario, the three decision makings could be very different, because the alternative phone models to choose from at the three time points $t_1$, $t_2$, and $t_3$ will be very different, as well as the popularity of the alternative models at these time points. The same applies to other products that have relatively a short life span on sales (from release to discontinued dates). In short, even if two users interact with the same item, if the two interactions occur at very different time points, the contexts for the two decision makings could be very different. The context here is reflected by the candidate items and their properties (\eg their popularity ranking) at the ``decision making'' time.

The context here could be just one of many factors that may affect user decision making. Modeling decision making or user preference from user behavior (\eg user-item interactions) is a complex issue.  We refer readers to~\cite{Jameson2022humanfactor} for a comprehensive discussion of human decision making for recommender system. Furthermore, \citet{WhatUsersWant22EC} also examine the inconsistency between user behavior and user true preference, and state that the inconsistency could be a reason for poor recommendations because the platforms are not optimizing for user happiness. Specifically, the authors formalize two decision-making agents for a user to model the inconsistencies of user behavior. Once decision making is ``impulsive and myopic'' (\eg enjoy watching shot-form videos for now) while the other decision is ``forward-looking and thoughtful'' (\eg shall not spend too much time watching videos).

In our discussion, we solely focus on the context changes at the system side, more specifically, the changes that can be fully observed in an offline dataset.\footnote{We do not consider context change at user side (\eg job change or relocation to a different place), or changes that cannot be observed in a dataset like a promotion campaign by a brand or seller.} For example, new items are released from time to time, and outdated items are no longer receiving more interactions from any user after some time. Other than availability and unavailability of items, different items accumulate different number of interactions at different time points.

Back to the illustration in Figure~\ref{fig:timeDecisionContext}, if user $u_4$ purchases the same phone within the few days of $u_3$'s purchase, then likely the contexts for these two decision makings are very similar. Although it is hard to directly model the context changes between any two user-item interactions, it is reasonable to assume that if two interactions occur within a short time period, the context change at system side is not significant. The length of a reasonable ``short time period'' may vary from one system to another depending on the characteristics of the items (\eg news, movie, music, book, restaurant, and consumer electronics)~\cite{Perspective22forgetting}. In other words, if a user has many interactions that occurred not too long ago from the test time, the contexts of the past interactions and the context of this test instance are similar. In this case, the test interaction may well align with the user preference learned from the past few interactions. This could be an explanation to why recommender models that are trained with  recent parts of the dataset deliver better accuracy~\cite{Perspective22forgetting}, and to why the users who have recent transactions enjoy better recommendations~\cite{ICTIR22Loyal}. 

Because of the ignorance of timeline in our modeling, the possible context changes cannot be considered in mainstream  RecSys models.  Two user-item interactions that occurred 10 years apart are modeled in the same way as if the two interactions occurred within the same day. This could be an implication of ignoring global timeline in our evaluation.

Here comes another question. If global timeline is so important, then \textit{why industry players are not highlighting this factor in modeling?} One possible reason is that industry players often need to process a large volume of data.  In their model evaluation (regardless of online or offline), the recent parts of data are sufficient for training. We can take some recently released datasets as examples. MIND dataset for news recommendation from Microsoft contains interactions of one million users randomly sampled in \textit{6 weeks}~\cite{MindDatasetMS}. The user behavior data from Taobao for recommendation contains interactions of one million users randomly sampled in about \textit{one week}.\footnote{\url{https://tianchi.aliyun.com/dataset/dataDetail?dataId=649&userId=1&lang=en-us}} In terms of dataset size, these two datasets are not small;  in terms of the time duration they cover, the system side contexts may not change much.\footnote{It is understandable that news recommendation is very sensitive to time due to the characteristics of the items.} In industrial-scale systems, \citet{MLAds22} state that ``limiting training data to more recent periods is intuitive.'' The authors further comment that if the date range is extended further back in time, ``the data becomes less relevant to future problems''. In industry setting, recommenders are often periodically retrained/updated with the recent data. For example, in the implementation of the Wide \& Deep learning which powers the Google Play recommendation,  \citet{WideDeep16} state that ``user and app impression data within a period of time are used to generate training data'', then ``every time a new set of training data arrives, the model needs to be re-trained''. A more recent model is Monolith~\cite{Monolith22Byte}, a BytePlus Recommend product. The training of Monolith is designed to have batch training stage and online training stage. For online training, the model parameters can be updated at minute-level; hence the model is able to ``interactively adapt itself according to a user’s feedback in realtime''~\cite{Monolith22Byte}. In this sense, practical recommenders naturally follow timeline and also consider data recency along timeline.

Recall that the ``the goal of the offline experiments is to filter out inappropriate approaches, leaving a relatively small set of candidate algorithms to be tested'' online~\cite{Gunawardana2022}. Evaluating our models in a manner that closely mimics their online usage remains of utmost importance. Doing so will help to narrow the divide between the algorithms described in academic papers and the ones that drive various platforms.

\section{A Fresh Look at RecSys}
\label{sec:freshlook}

In academic research, we often abstract similar real-world problems (\eg the various types of recommendation problems in different domains) to a formal research problem. Accordingly, we propose evaluation metrics to quantify to what extent a proposed solution has addressed this problem on different datasets. Very often, a solution is designed largely for the purpose of achieving better evaluation scores. 

Unfortunately, the abstraction process to reach a formal  problem definition comes along with simplification of  real-world problems. In the mainstream RecSys problem setting, the ``time'' factor has been largely ignored due to the simplification process.  Recall the five settings for offline evaluation  in Table~\ref{tab:5settings}. Setting 1 is the most close to the real-world scenario  while Setting 5  completely ignores timestamps. Our case study shows that 59.1\% of papers adopt Settings 4 and 5, which is a strong indication of problem simplification. 

To give a fresh look at the RecSys problem, the global timeline has to be part of the problem definition, to truly reflect the problem of learning from \textit{past} interactions then to recommend \textit{unseen} items. Accordingly, the evaluation methodology has to factor in the global timeline.

\subsection{More Practical Evaluation}

We are not in short of papers on large-scale empirical evaluations~\cite{rigorousEvaluation,futureItemRec,worrying,DaisyRec22,Bars22Evaluation}. However, the results reported in these papers may not be comparable to each other due to the ignorance of timeline. A recent benchmark~\cite{futureItemRec} verifies that current evaluation methodology leads to recommending ``future items'' which would never occur in reality, a consistent finding earlier reported in~\cite{dataLeakage}. Despite the existence of  many large-scale empirical studies, there remain questions on reproducibility, and technical and theoretical flaws~\cite{Crisis21AIM,worrying21tois}. On the other hand, it is a challenging problem to evaluate recommender systems, because the evaluation metrics can be defined from multiple perspectives~\cite{RecSysEveSurvey22,AIM22OfflineChallenge,Jannach_Bauer_2020} and there remain many challenging issues even if these metrics are well defined~\cite{MetricConsistent21}, and even if the evaluation is conducted online~\cite{commonPitfalls}. 

We probably want to begin with something simple: \textit{Can we fairly compare our model with Popularity, the simplest baseline?} In other words, our recommenders shall be compared with a Popularity ranking that is used in practice, \ie the ranking of items on a hourly, daily, or weekly basis.

\begin{figure}
    \centering
    	\includegraphics[trim = 4.1cm 12.6cm 15.1cm 1.2cm, clip, width=0.95\columnwidth]{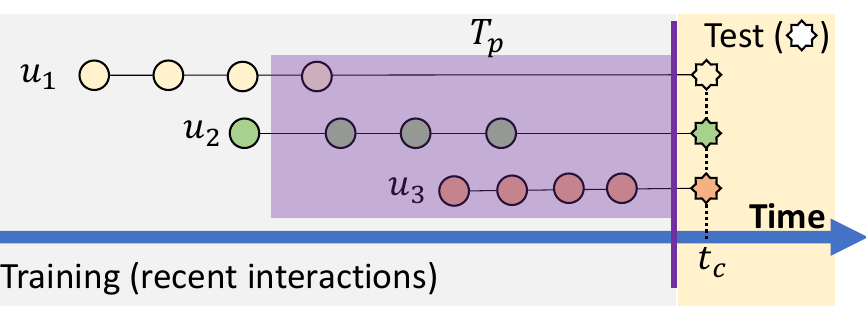}
      \caption{Illustration of evaluation of (recent) popularity along timeline. The popularity of an item is derived based on the interactions received in time duration $T_p$.}
          \label{fig:timePopEva}
	\Description{}
\end{figure}

In Figure~\ref{fig:timePopEva}, the vertical line in purple  is an illustration of Popularity ranking of items at a time point slightly before the current time $t_c$. When users interact with the system at $t_c$, the popularity ranking provides the most popular items in the past $T_p$ time duration. $T_p$ can be one hour, one day, one month, or the entire history, \ie a parameter depending on the characteristics of the items. This popularity method can be more precisely named as \textit{Recent Popularity} where the \textit{recency} $T_p$ is configurable. If $T_p$ is set to cover the entire duration of all existing training data, then the ranking is the most popular items in history. Note that the most popular in history remains different from the ill-defined Popularity baseline, because of the observation of the global timeline in Figure~\ref{fig:timePopEva}.

\begin{figure}
    	\centering
    	\includegraphics[trim = 4.3cm 10.8cm 15.1cm 3cm, clip, width=0.95\columnwidth]{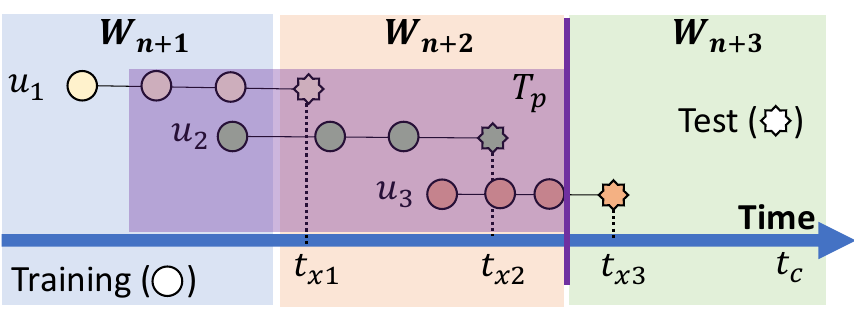}
    \caption{Illustration of the timeline scheme. Models are evaluated window by
window, and each time all test instances within one time window are evaluated. }
    \label{fig:evaTimeline}
	\Description{}
\end{figure}

\subsubsection{The Timeline Scheme.} 
We now apply a similar concept to evaluate any RecSys model in offline setting. The first step in the evaluation is to split  data into training and test sets. We may follow any existing data split scheme. In the following, the \textit{leave-one-out} scheme is used as an example. With leave-one-out split, the last interaction of every user is in the test set and the remaining interactions of the user are in the training set. 

To observe the global timeline, all user-item interactions (in both train and test) are arranged in chronological order by their timestamps. Then the entire timeline is split into time windows of size $W$ as shown in Figure~\ref{fig:evaTimeline}. 
We evaluate a model on all test instances within one time window $W$ at each time, window by window. Suppose $W_{n+2}$ is the current window for evaluation, within which $t_{x1}$ and $t_{x2}$ are the two test instances (see Figure~\ref{fig:evaTimeline}). The model shall be able to learn from \textit{all or subset of} the following data instances: (i) all training instances in $W_{n+2}$, and (ii) all training and test instances in all the windows before $W_{n+2}$. For each test instance, we compute the evaluation measures (\eg precision/recall), and aggregate the results. The aggregation may happen at each window, or across all test instances in all windows. 

The proposed timeline scheme can be considered as a scheme sitting between Setting 2 and Setting 3 if mapping to the 5 settings in Table~\ref{tab:5settings}. It offers evaluations at multiple time points (through time windows) and maintain global timeline.  Note that, Setting 3 in Table~\ref{tab:5settings} (or single time split) evaluates a model at a single time point only. 

\subsubsection{Discussion on the Timeline Scheme.} 
We note the following points for the timeline evaluation scheme described above.

First, this evaluation scheme observes global timeline by design. Second, this evaluation scheme still suffers from data leakage. However, the amount of data leakage is  controlled by the size of time window $W$.  If the time window is reasonably small with respect to the entire timeline (\eg one month vs ten years), then data leakage may not significantly affect the results. The amount of data leakage in each time window trades off the total number of windows in the entire evaluation. We may consider that the current mainstream offline evaluation is a special case of the timeline scheme:  there is only one big time window covering the entire timeline duration. To completely avoid data leakage, the evaluation scheme can be changed to only allow the model to learn from previous windows when evaluating on the current time window.  Third, along timeline, the model needs to be retained or updated for evaluation on each time window. The design allows a model to learn from different sets of training instances. For example, recency popularity may derive the popularity ranking from only one or a few recent windows. Other machine learning based models may choose to use training instances from recent windows as well, \ie only the recent parts of the data. In this sense, the number of recent windows/interactions becomes a hyper-parameter in model training.  This is a different understanding of ``fair comparison'' from the current mainstream setting, where all models use the same set of training data.

Lastly, the timeline scheme is just one possible way to evaluate RecSys in a more practical manner and the scheme might have already been used in some previous studies. In particular, \citet{TemporalCF2009} conducted experiments on Netflix data with a window size of 7-day and evaluated two recommenders with dynamic modal updating along time. There are other data partition schemes that do not lead to data leakage, for example, partition by time point. Further, there is also a line of research in RecSys known as incremental learning. In this setting, a model learns from past data and predicts future interactions along timeline, by definition. The discussion here is to offer a revisit to the batch-oriented train/test offline evaluations that do not consider timeline.

The key consideration here is to observe global timeline in the evaluation process. The key assumption here is that the system context does not change much within a time window. As aforementioned, this evaluation scheme is not problem free, \eg data leakage remains possible. However, exactly mimicking real-time setting (\ie Setting 1 in Table~\ref{tab:5settings}) would lead to a very complex evaluation process, which is less practical in academic research. The timeline scheme described above is a relaxed version of Settings 1 and 2. 

A potential challenge of observing timeline in evaluation is data sparsity in RecSys. With timeline in consideration, the cold-start issue becomes a common problem for almost every user. Basically, with time in the picture, the user-item interactions are no longer projected to a two-dimensional space (\ie user and item), but a three-dimensional space (\ie user, item, and time). This would make the RecSys problem much more complicated and challenging, but could be more interesting as well. At the same time, data sparsity could be partially eased if additional information (\eg attributes) are available with users and/or items in the released dataset. This also calls for a revisit of dataset release from industry on what information can be released in addition to the user-item interactions, without compromising user privacy.

\subsection{Meaningful User Preference Modeling}

A recommender is expected to answer a user's latent (information) needs.  The user-item interactions are the observed results, or the answers to the earlier information needs. In the original design of collaborative filtering, users choose who to follow as the expected ``information filters''. In the current RecSys model design, users are not empowered to proactively choose the information filters. Rather, the information filters are modeled based on users past interactions. However, such simplification in user preference modeling only captures the results of the decision makings, and does not capture the ``contexts'' of the decision makings. Even if two users interact with the same item, they may not make decisions in a similar context, particularly when the two decision makings occur a long time apart from each other. However, it is hard to model the decision context, given the limited data in academic research. In the following, we discuss a few possible ways to model context similarity between two decision makings. Nevertheless, more and deeper research is expected in this area considering decision making is a complex issue~\cite{Jameson2022humanfactor,WhatUsersWant22EC}.

One possible way of evaluating similarity of decision contexts is through impressions. In simple words, impressions are a list of items presented to a user when she/he makes an interaction. For example, if user $u_1$ interacts with item $D$ when presented with impression $\{A, B, C, D\}$, and user $u_2$ also interacts with item $D$ with impression $\{D, E, F, G\}$, then the two decision makings are based on different contexts, although the final decisions are the same. Recently, a few datasets are made available with impressions, including ContentWise Impressions~\cite{ContentWise20}, MIND~\cite{MindDatasetMS}, and FINN.No Slates~\cite{FINNDataset} datasets. Such datasets are of great value for exploring new ways of user preference modeling. There is also a study on evaluation of recommender systems with impressions~\cite{Evaluation22Impression}.   

Large-scale recommender systems typically consist of multiple steps like matching  (\ie candidates generation) and ranking~\cite{Bars22Evaluation,Reranking22Survey}. In the matching step, candidate items are identified from all available items. In the absence of impressions, the decision context might be reflected by these candidate items, provided that the matching step observes the global timeline and retrieves only the available items at the corresponding time point (\ie a test instance's timestamp). Here, the similarity of candidate items is a proxy to measure the similarity of two decision contexts. The reason of using  candidate items instead of the final ranked items is that  candidate items are less dependent on a particular ranking algorithm.  To generate candidate items for every interaction is very expensive. A simplification here could be an assumption that if two interactions happen within a very short time period, then the decision contexts  are similar. Then, a function of time duration could be used to model the context changes between two interactions.

Modeling user preference with the consideration of decision context offers us a fresh look at many specific problems in RecSys. One of them is sequential recommendation. In sequential recommendation, interactions (or actions) of a user form a sequence and the task is to predict the user's next interaction. A user sequence preserves the relative sequential order of her interactions, but does not record the timestamps of these interactions. \citet{SeqRecencySampling22} show that recent training interactions (in terms of sequential order)  in a sequence  better indicate the user's interest. The recency by sequential order can be an approximation of the recency by timestamps. 
Nevertheless, without recording  timestamps of the interactions in a sequence, it remains difficult to precisely model the context differences in decision making. For one user, her first and last interactions could be one year apart. For another user, all interactions in her sequence may occur within a week. The decision making contexts in these two sequences will be quite different. Hence, it would be more meaningful to model interactions with timestamps in sequences.  By considering timestamps, we are also in a better position to evaluate whether some datasets are indeed suitable for sequential recommendation. For example, the MovieLens dataset is not suitable for sequential recommendation, because there is no meaningful sequence in the dataset~\cite{pseudoSequence}. 

\section{Discussion}
\label{sec:discussion}

First, considering temporal factor in offline evaluation is not new, evidenced by the 5 settings listed in Table~\ref{tab:5settings} originated from~\cite{Gunawardana2022} and its earlier edition~\cite{GunawardanaS2015}. An example evaluation with a sliding window has also been reported in 2009~\cite{TemporalCF2009}. Nevertheless, as shown in Table~\ref{tab:datasplit}, the mainstream offline evaluations do not consider global timeline. Simulation-based online settings are being increasingly adopted in recent studies primarily because of the algorithms they use, such as reinforcement learning, rather than a general need for recommendation models. Researchers who are new to this field may follow the bulk of existing literature and simply borrow the widely adopted offline evaluation settings without much further consideration. This paper offers an alternative view for researchers to reconsider how to conduct offline evaluations to better serve the purpose of selecting best candidate algorithms. 

Second, this paper tries to provide an alternative view of ``fair comparison''. If two methods $A$ and $B$ are given access to the same set of training data, then it becomes a method's own choice on which portion and how to use the training data. For example, if a Popularity method achieves its best performance by using a weekly updated ranking along time, then it should not be forced to use all historical data to produce a static ranking. Similarly, if a machine learning model performs the best by learning from only the recent portion of the training data, then it is fair to compare with another model which learns from more historical data, as long as both models have access to the same training set. In this case, the amount of training data to use becomes a model hyperparameter as studied in~\cite{Perspective22forgetting}. 

Third, as a new look at recommender systems, this paper proposes a timeline scheme for offline evaluation as a  better way to simulate a model's practical setting. Again, this general concept of the timeline scheme is not new. However, the key message here is to maintain global timeline in evaluation when the dataset covers a long time period. The scheme is designed to balance the evaluation complexity (\eg the number of models need to be trained or updated along time) and the potential issues (\eg data leakage). On the other hand, a potential risk here is that once ``time'' becomes a parameter in the evaluation process (\eg the size of time window in the timeline scheme), it may become yet another parameter to be optimized for through system design. More research is expected to come out with a more effective and accurate evaluation mechanism for offline RecSys evaluation.

Fourth, this paper also briefly touches the concept of modeling decision making contexts, rather than the results of decision making (\ie user-item interactions) for user preference modeling. To our understanding, decision making context is time-dependent. With the timeline scheme in evaluation and the models are retrained/updated along timeline, the modeling of decision contexts could lead to more interesting findings in recommender systems. 

Lastly,  it should be noted that the criticisms raised in this paper regarding poorly executed evaluations are primarily directed towards recommender systems in academic research. While industry practitioners typically use  recent data to train and evaluate their models and regularly update them, we believe that there is a pressing need to improve the quality of offline evaluations in academic research in order to bridge the gap between academic and industry practices. On the other hand, the extent to which our conclusions about the model effectiveness would be altered by taking the global timeline into account during evaluation has not been thoroughly investigated.

\section{Conclusion}
\label{sec:conclude}
We start with a few counter-intuitive observations made in recent studies; then we explain the reasons behind. One key reason is the ignorance of the global timeline in model evaluation, which leads to a poor implementation of the simplest baseline Popularity. Interestingly, because industry players often have to limit to recent data due to large data size, they may not highlight the importance of timeline. However, in academic research, many widely used datasets cover interactions recorded in a long period. Following a similar problem understanding as industry players but evaluating on datasets of different characteristics is the main contradiction here. The missing of timeline leads to improper evaluation of our models as the offline settings are far away from a good simulation of the online scenario. Hence, the models developed in academic research are rarely transferable to practical systems. In this paper, we highlight the importance of timeline for a better simulation of the online setting and hence more indicative results of which algorithms are more worth the expensive online testing. On top of that, the consideration of timeline also provides us new insights in modeling user preference. We shall not only focus on the user-item interactions, which are the results of decision makings, but also focus on the contexts of decision makings. After all, we aim to model user preference in making decisions. 

This perspectives paper calls for research in the following directions. One is the adaptation and standardization of the timeline evaluation scheme, where global timeline is build in the evaluation process. This direction includes two subtasks: (i) the study on the extent to which our existing conclusions about model effectiveness would be altered by taking the global timeline into account, and (ii) the complexity of introducing time as another factor in the evaluation process. Another direction is the deep understanding of the relationship between user behavior and recommendation. This direction also includes two subtasks: (i) a better way of interpreting user-item interactions as the results of decision makings for effective recommendation, and (ii) a better way to understand and model user decision making.

\begin{acks}
The author would like to thank Chunyang Wang for data preparation, and thank Yitong Ji, Shuai Zhang, Jie Zou, Xin Zou and the anonymous reviewers for the constructive feedback and insightful comments.  
\end{acks}

\bibliographystyle{ACM-Reference-Format}
\balance
\bibliography{RecSysNewLook}


\end{document}